*Research Article*

# Symmetry and the Arrow of Time in Theoretical Black Hole Astrophysics

**David Garofalo**

*Department of Physics, Kennesaw State University, Marietta, GA 30060, USA*

Correspondence should be addressed to David Garofalo; dgarofal@kennesaw.edu

While the basic laws of physics seem time-reversal invariant, our understanding of the apparent irreversibility of the macroscopic world is well grounded in the notion of entropy. Because astrophysics deals with the largest structures in the Universe, one expects evidence there for the most pronounced entropic arrow of time. However, in recent theoretical astrophysics work it appears possible to identify constructs with time-reversal symmetry, which is puzzling in the large-scale realm especially because it involves the engines of powerful outflows in active galactic nuclei which deal with macroscopic constituents such as accretion disks, magnetic fields, and black holes. Nonetheless, the underlying theoretical structure from which this accreting black hole framework emerges displays a time-symmetric harmonic behavior, a feature reminiscent of basic and simple laws of physics. While we may expect such behavior for classical black holes due to their simplicity, manifestations of such symmetry on the scale of galaxies, instead, surprise. In fact, we identify a parallel between the astrophysical tug-of-war between accretion disks and jets in this model and the time symmetry-breaking of a simple overdamped harmonic oscillator. The validity of these theoretical ideas in combination with this unexpected parallel suggests that black holes are more influential in astrophysics than currently recognized and that black hole astrophysics is a more fundamental discipline.

## 1. Introduction

The laws of nature appear to be time-reversal invariant but when they operate in the arena of our messy astrophysical Universe filled with gas, dust, and stars, friction destroys the time symmetry of the solutions and a past/future asymmetry is produced. The more complex the frictional damping present in the Universe is, the more the underlying time symmetry gets washed away. Because astrophysics is grounded in the time evolution of radiation and matter fields, such symmetries are ignored. But the tendency to ignore time-reversal symmetry in this discipline amounts to assuming that the symmetry is well hidden, which, in turn, implies that astrophysics is a less fundamental branch of theoretical physics, that is, that the large-scale processes at work in the astrophysical Universe will shed little light on these fundamental theoretical ideas. But the degree to which this is true has not been explored.

The so-called "gap paradigm" for black hole accretion, jet formation, and active galaxy evolution constitutes a phenomenological framework unifying active galaxies based on the change in size of the gap region [1]. Due to the one-parameter dependence of this theory, it appears to be a strongly coarse-grained framework whose validity argues that complicated or chaotic astrophysical processes matter less. But this is surprising for a theoretical structure that attempts to explain the complicated dynamical properties of hot, magnetized, accreting matter onto black holes. In fact, black hole accretion appears to require algorithms with a sophistication that is beyond our current numerical magnetohydrodynamical (MHD) capabilities (i.e., including nonideal MHD effects, causal Ohm's law, and radiation). It would seem, thus, that delving further into the complications of nonlinear MHD is the only way to go and that attempts at simplifying the dynamics are bound to provide us with only a superficial understanding. Therefore, a one-parameter model that captures anything interesting in black hole astrophysics is puzzling.

But simplicity reminds us of basic physics. The simple harmonic oscillator, for example, is characterized by solutions to the equations of motion displaying time-reversal symmetry, a symmetry that seems to live at the foundations of all the basic laws of physics. Our goal here is to argue that the puzzling validity of this gap paradigm has something to do not only with the fact that black holes are simple objects and thus should at some level display a time-reversal symmetry, but also with the fact that their influence is sufficiently dominant that evidence of this symmetry is actually visible in the distribution of active galaxies over time. We argue this by highlighting the presence of symmetries in the model for active galactic nucleus energy generation that are reminiscent of those in the basic laws of physics, suggesting that the presence of such symmetries in black hole astrophysics is a sign of the tight coupling that exists between black holes and the large-scale host galaxy. By analogy with the damped harmonic oscillator, we describe how accretion and jets in the gap paradigm appear to conspire to break the time-reversal symmetry in the theory but in a mild enough form that the symmetry remains visible. We accomplish this by illustrating how the character of the so-called jet-disk connection—a currently unsolved and strongly debated issue in astrophysics—subsumes the role of friction in damping the harmonic oscillator-like behavior that is evident in the basic physics of jet power. In Section 2 we provide an elementary description of time-reversal symmetry; in Section 3 we present our hybrid numerical/analytic solution upon which we show how the gap paradigm appears to incorporate the time-reversal symmetry; and, in Section 4, we conclude.

## 2. Time-Reversal Symmetry of the Basic Laws of Physics

Time-reversal symmetry is easily and straightforwardly described in terms of the classic, nonrelativistic, simple harmonic oscillator, whose world line is plotted in Figure 1. By adopting the origin as initial condition, the symmetry is directly manifest; namely, there is degeneracy in the choice of positive or negative times in modeling the evolution of the oscillator. The dynamics is toward equilibrium either way. Hence, the solution displays time-reversal symmetry. More generally, and without reference to initial conditions, describing the dynamical evolution of the oscillator can be accomplished by adopting either positive or negative time. But, by introducing damped motion, we can break the symmetry in time (Figure 2). Since the energy of the oscillator is proportional to the square of the amplitude, this means that time symmetry-breaking is equivalent to a damping or an energy loss on the part of the oscillator.

Hence, friction produces the past/future asymmetry familiar from direct experience. Note that although the symmetry can be broken in ways that make it difficult to recognize or reconstruct from the complications of damped world lines, it is there nonetheless. And, of course, the more dramatic and complex the nature and magnitude of the damper, the more complex the task of identifying the original symmetry. Since astrophysics deals with the myriad complexity of different types of matter in the Universe, the whole shebang of scratching boards, so to speak, one might expect the real world to produce a complex enough damping that any symmetry becomes difficult or impossible to identify. But this is simply a well-motivated assumption. In the next section we describe the basic elements of the gap paradigm, highlighting the visibility of both the apparent damping and the underlying symmetry. The point of this, again, is to illustrate a surprising connection between the properties and evolution of matter on large scales and the symmetries in the basic laws of physics that appear to be reflected in the behavior of matter on large scales, that is, that fundamentally simple physical systems, such as classical black holes, leave traces of their simplicity on large scales.

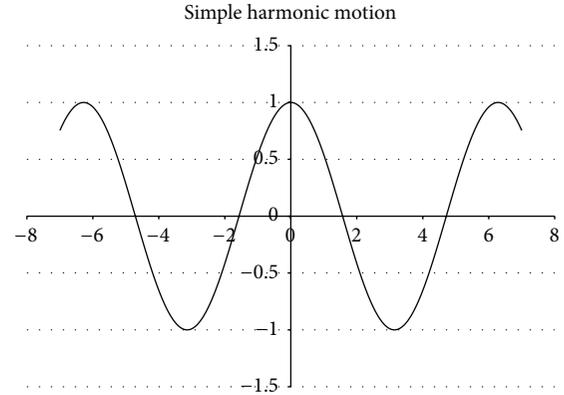

Figure 1: Position versus time for the simple harmonic oscillator.

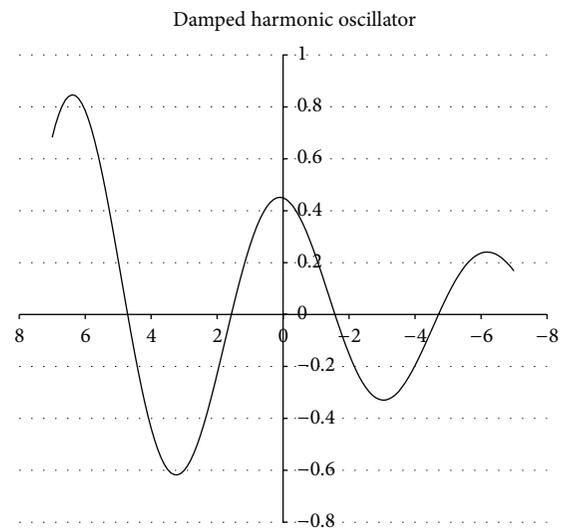

Figure 2: Damped harmonic motion with the amplitude decreasing in time illustrating the time asymmetry of a Universe with friction.

## 3. Symmetry-Breaking in Black Hole Astrophysics: The Gap Paradigm

The gap paradigm combines three independent theoretical constructs into one global phenomenological model. The most fundamental one involves the physics of energy extraction from black holes via the Blandford-Znajek effect [2], which postulates the relation as follows:

$$L \propto a^2 \qquad (1)$$

between extracted power and black hole spin parameter $a$. The other two constructs involve extraction of accretion disk rotational energy via Blandford-Payne jets [3] and accretion disk winds ([4, 5] as extensions of [6, 7]). The total outflow power from the Blandford-Znajek effect, the Blandford-Payne mechanism, and the Kuncic-Bicknell disk wind is based on the size of the gap region which is imposed on the following standard set of equations. The magnetosphere is governed by the standard Maxwell equations with sources, which relate the exterior derivative of the dual Faraday 2-form to the current as follows:

$$dF_* = \mu_0 J \qquad (2)$$

plus the force-free condition on the Faraday 2-form as follows:

$$F \cdot J = 0 \qquad (3)$$

and, finally, we also impose the dissipationless ideal MHD condition:

$$F \cdot U = 0, \qquad (4)$$

where $U$ is the velocity field of the accretion flow. In the accretion disk, instead, the dissipationless condition does not apply and we relate the current to Ohm's law via

$$dF_* = \sigma F \cdot U \qquad (5)$$

with $\sigma$ the conductivity, which we treat as constant in both space and time. Since we work directly with the vector potential $A$, we have

$$F = dA \qquad (6)$$

and our equations take the following form:

$$dA_* = \mu_0 J,$$
$$dA \cdot J = 0, \qquad (7)$$
$$dA \cdot U = 0$$

in the magnetosphere, while the accretion flow is constrained by

$$d \wedge dA_* = \sigma dA \cdot U. \qquad (8)$$

Stationarity and axisymmetry fully constrain the gauge. We seek solutions of

$$\int \frac{F}{2\pi} = \Psi \qquad (9)$$

for a ring constructed using fixed radial and poloidal coordinates in Boyer-Lindquist coordinates. This flux function allows us to determine Blandford-Znajek and Blandford-Payne power [1]. On top of that we add the power associated with the disk wind which involves an integral over the entire accretion disk of the local dissipation function. This function can be obtained from Shakura and Sunyaev [6] to be

$$D(r) = \left(\frac{3}{8\pi r^3}\right) GM \left(\frac{dm}{dt}\right) \left[1 - \left(\frac{R_{\text{isco}}}{r}\right)^{1/2}\right]. \qquad (10)$$

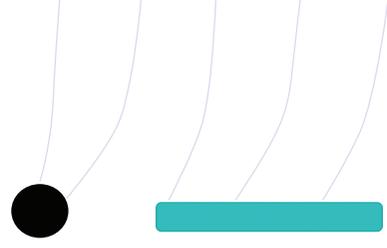

FIGURE 3: A black hole, a gap region, and an accretion disk in cross section.

$M$ is the mass of the black hole and $dm/dt$ is the accretion rate. Hence, the wind power at any location $r$ depends on the location of innermost stable circular orbit, $R_{\text{isco}}$. For locations further out in the disk, the local dissipation from the disk will be greater in the prograde configuration due to the smaller value of $R_{\text{isco}}$. This nonrelativistic calculation is sufficient given that the relativistic correction factors drop off rapidly with $r$.

In the teal region of Figure 3, as well as in the equatorial plane of the gap region, and on the black hole horizon, the plasma is dominated by the inward radial gravitational pull (with a slow radial infall velocity in the disk and rapid radial infall velocity in the gap region), while, everywhere else, the plasma is in a force-free state; that is, it flows on the field lines or more precisely on the contours of constant flux. As the gap region changes in size with spin, the three power generating mechanisms vary. While the two jet producing effects decrease in power as the gap region decreases in size, the disk wind power increases as can be discerned from the dissipation function dependence on $r$. We will show how the solutions to our equations produce a total outflow power that resembles the time evolution of a simple overdamped harmonic oscillator.

In Figure 4 we show the symmetric behavior of Blandford-Znajek power with respect to the spin parameter. Negative spin values represent an accretion disk counterrotating with respect to the black hole while positive spin values represent corotation between the accretion flow and black hole. But the gap paradigm postulates motion away from negative spin values for forward evolution in time and therefore towards positive values [1]. Thus, exchanging spin with time produces dynamical evolution. This is the crucial point because it is here that the invariance of the dynamics with respect to time-reversal enters the picture and the harmonic oscillator analogy becomes transparent. So how does the power evolve from $L \propto a^2$ relation to $L \propto t^2$, where $t$ is time? The answer is accretion. Accretion of mass onto the black hole spins the black hole up towards the angular momentum value of the accretion flow. If the system is in a retrograde configuration with the accreting material rotating opposite to the direction of the black hole angular momentum, the magnitude of the holes' angular momentum will decrease towards zero spin and then spin up in the prograde direction. In other words, a continuous accretion scenario allows one to replace spin with time. While accretion in principle leads to symmetric power dependence as a function of time between retrograde and prograde configurations, the actual physical accretion process introduces an asymmetry between the two regimes. In the retrograde case, in fact, not only is the angular momentum that is supplied to the black hole greater in magnitude (because the innermost stable circular orbit is further away compared to the prograde case), but the difference between the angular momentum of the black hole and accreted material is larger. Therefore, the time to spin down a maximally retrograde black hole at a fixed accretion rate is less in the retrograde regime by about a factor of 10. It is important here to recognize the symmetry-breaking produced by accretion. If the accreting matter organized itself in such a way to produce a symmetric feeding of angular momentum to the black hole during retrograde and prograde regimes, the power versus time would be time-reversal symmetric. The jet power, in fact, depends on the magnitude of the magnetic field threading the black hole and the magnitude of its spin so the black hole energy extraction mechanism by itself does not produce any time asymmetries. The point is that mechanisms that are external to the black hole are responsible for breaking the degeneracy of the underlying time symmetry. Note that a different even function spin dependence is compatible with this analysis; that is, the argument is not restricted to $L \propto a^2$. There is no fundamental expectation that the power dependence should involve the spin squared. The origin of that derivation is phenomenological and numerical simulations have suggested other even spin dependencies.

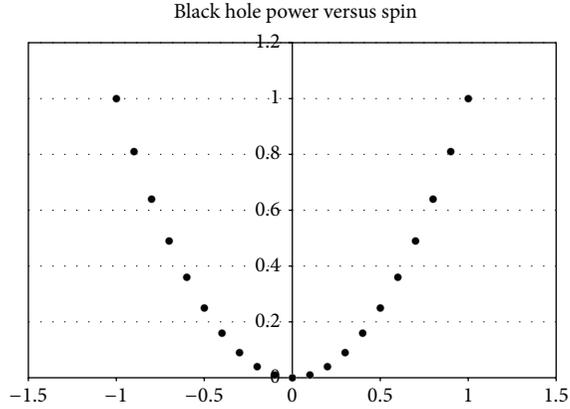

FIGURE 4: Black hole power (arbitrary units) versus dimensionless spin showing the symmetry of the underlying black hole energy extraction mechanism.

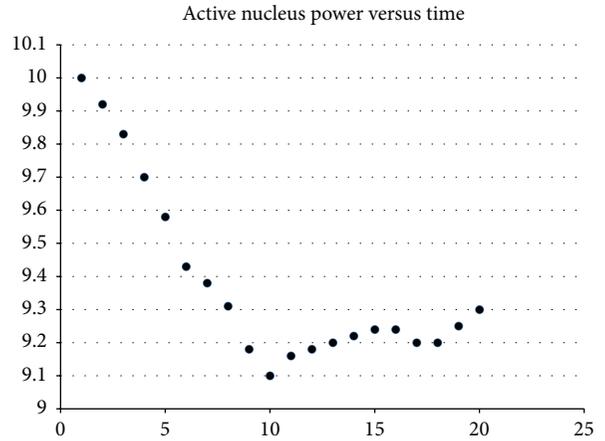

FIGURE 5: Total power including all three mechanisms of the gap paradigm (arbitrary units) emitted from the active nucleus as a function of time in units of $10^7$ years. Values are determined as an average over a wide range of AGN as prescribed in the paradigm in the sense that the outflow features are a function of the gap region.

When Blandford-Payne disk jets and Kuncic-Bicknell disk winds are included, their combined effect breaks the symmetric behavior of Blandford-Znajek by increasing the amplitude toward the negative spin values and decreasing it in the positive direction as shown in Figure 5. As described in Garofalo et al. [1], powerful FRII jets, modeled as retrograde black hole accretion systems, heat and expand the interstellar medium leading to a phase of hot Bondi accretion in which the accretion disk is no longer radiatively efficient. If, instead, the jet is less powerful and less efficient in heating the surrounding medium, accretion remains longer in its original radiatively efficient mode. But time evolution toward positive spin values means accretion disks live closer to black holes and their effect is to smother the jet. This tug-of-war determines whether the active galaxy evolves toward a less powerful radio loud system (or FRI jet) or radio quiet one. Either way, both jet power and accretion power decay with time. Hence, positive time witnesses the required damping but in a mild enough form that the underlying symmetry remains visible. Each data point in Figure 5 is an average over the different kinds of AGN weighted by the observed fraction of radio loud versus radio quiet systems at about 20%. FRII radio galaxies and FRII quasars contribute a smaller fraction of the total disk power due to their larger gap regions while radio quiet quasars contribute the most since they have the smallest gap regions. But in time the fraction of objects with large gap regions decreases and the jet power drops. However, for objects that evolve into radiatively inefficient accretion modes, their prograde spins are again associated with powerful jets that are weaker than in the retrograde regime by an order of magnitude or so depending on the exact spin value [1]. The evolution to the prograde regime and the decrease in accretion ensures that the curve fails to make it to the peak power reached at early times, in qualitative accord with the observed AGN luminosity function and the damped oscillator.

Nothing fundamental, surprising, or even worth noting emerges from the recognition of a general decay. That is just the second law of thermodynamics. But the fact that so much of that decay can be traced back to black hole details is worth emphasis. In other words, energetic processes are expected to decay in time as their complex dynamics satisfy the second law of thermodynamics. But, again, one expects the complex nature of astrophysical processes to reveal or reflect little of the underlying simplicity of the basic physical processes. Yet, in this model, Figure 5 is both a statement of the time evolution of active galaxies as a whole and of the black hole engine, two macroscopic realms separated by an enormous gap in scales that in turn seem to share the same simple damped oscillator description. Notice that the power generated by the accretion disk is

$$L = \left(\frac{GM}{2R}\right) \frac{dm}{dt},\quad(11)$$

where $M$ is the mass of the black hole, $R$ is its radius, and $dm/dt$ is the accretion rate. Producing time-reversal symmetry in the above expression requires the symmetry to appear in the accretion rate. But no such expectation exists, of course, as accretion rates simply die down with time as the available reservoir of matter drops. But even if we impose a constant source of matter and assume a constant accretion rate, the conversion of gravitational potential energy into radiation and magnetic fields is an inherently irreversible process. The expectation, more generally, is that any astrophysical process—not only accretion—should display this irreversibility. While the symmetry in the black hole engine itself may not completely surprise, observational evidence of such symmetries in the distribution of active galaxies should. And this work is an attempt to highlight that fact. In other words, from the perspective of the gap paradigm, the observational fact of a distribution of active galaxies with FRII quasars at highest average redshift, FRI radio galaxies at lower average redshift, and smaller-sized radio quiet accreting black holes at lowest redshift reflects the simple, visible, symmetry-breaking of the black hole engine.

## 4. Conclusions

This work identifies a time symmetry in the black hole power extraction of the gap paradigm, tempered by the presence of accretion. While the symmetric component is associated with magnetic fields threading the black hole itself, the damping mechanisms of accretion are, of course, external to the black hole. The point, however, is that they operate on distance scales that are on the order of a few tens of gravitational radii or less which means they are mechanisms that are provincial to the black hole, very much associated with the activity of the black hole itself and not with the large-scale galaxy. However, despite damping mechanisms that locally influence the black hole environment and that one might expect would quickly produce large gradients in entropy, the symmetry of the black hole surprisingly appears to survive not only locally but on large scales in the distribution of active galaxies as a whole. This generates a twofold suggestion, one in black hole astrophysics which is that black holes matter more to galaxies than currently appreciated and the other in theoretical physics which is that as a discipline black hole astrophysics is more fundamental than heretofore realized.

## Conflict of Interests

The author declares that there is no conflict of interests regarding the publication of this paper.